\documentclass{icrc2009}

\usepackage{graphicx}   
\usepackage{caption}    
\usepackage[font=footnotesize]{subfig} 
\usepackage{fixltx2e}
\usepackage{url}
\usepackage{amssymb}
\usepackage{lineno}

\newcommand{\shorttitle}[1]%
{\markboth{Proceedings of the 31\MakeLowercase{$^{st}$} ICRC, {\L}\'{o}d\'{z} 2009}{#1} }
\newcommand{\etal}{\MakeLowercase{\textit{et al. }}} 

\begin{document}
\title{JEM-EUSO Science Objectives}

\author{\IEEEauthorblockN{Medina-Tanco G.\IEEEauthorrefmark{1},
	Asano K.\IEEEauthorrefmark{2},
	Cline D.\IEEEauthorrefmark{3},
	Ebisuzaki T.\IEEEauthorrefmark{4},
	Inoue S.\IEEEauthorrefmark{5},\\
	Lipari P.\IEEEauthorrefmark{6},
	Parizot E.\IEEEauthorrefmark{7},
	Santangelo A.\IEEEauthorrefmark{8},
	Sigl G.\IEEEauthorrefmark{9},
	Takahashi Y.\IEEEauthorrefmark{10},
	Takami H.\IEEEauthorrefmark{11},\\
	Teshima M.\IEEEauthorrefmark{12},
	Weiler T. J.\IEEEauthorrefmark{13},
	for the JEM-EUSO Collaboration}
                            \\
\IEEEauthorblockA{\IEEEauthorrefmark{1}Instituto de Ciencias Nucleares (ICN), Universidad Nacional Aut\'onoma de M\'exico, M\'exico D.F.}
\IEEEauthorblockA{\IEEEauthorrefmark{2} Interactive Research Center of Science, Graduate School of Science,
Tokyo Institute of Technology, \\
2-12-1 Ookayama Meguro-ku Tokyo 152-8550, Japan.}
\IEEEauthorblockA{\IEEEauthorrefmark{3} Department of Physics and Astronomy, University of California, Los Angles, USA.}
\IEEEauthorblockA{\IEEEauthorrefmark{4} RIKEN Advanced Science Institute, Japan.}
\IEEEauthorblockA{\IEEEauthorrefmark{5} Dept. of Physics, Kyoto University, Kyoto 606-8502, Japan.}
\IEEEauthorblockA{\IEEEauthorrefmark{6} INFN-Roma La Sapienza, I-00185 Roma, Italy.}
\IEEEauthorblockA{\IEEEauthorrefmark{7} APC - AstroParticule et Cosmologie, CNRS : UMR7164 - CEA - IN2P3 - Observatoire de Paris,\\ Universit\'e Denis Diderot - Paris VII, France.}
\IEEEauthorblockA{\IEEEauthorrefmark{8} Institute fuer Astronomie und Astrophysik Kepler Center for Astro and Particle Physics\\ Eberhard Karls University Tuebingen Germany.}
\IEEEauthorblockA{\IEEEauthorrefmark{9} Institut theoretische Physik Universitaet Hamburg Luruper Chaussee 149 D-22761 Hamburg, Germany.}
\IEEEauthorblockA{\IEEEauthorrefmark{10} Dept. of Physics, The University of Alabama in Huntsville, Huntsville, AL35899, USA}
\IEEEauthorblockA{\IEEEauthorrefmark{11} Institute of the Physics and Mathematics of the Universe, the University of Tokyo, 5-1-5, Kashiwanoha,\\ Kashiwa, Chiba 277-8568, Japan}
\IEEEauthorblockA{\IEEEauthorrefmark{12} Max-Planck-Institut f\"ur Physik, F\"ohringer Ring 6, D-80805 M¬unchen, Germany.}
\IEEEauthorblockA{\IEEEauthorrefmark{13} Department of Physics and Astronomy, Vanderbilt University Nashville, TN 37235 USA.}
}

\shorttitle{Medina-Tanco G. \etal JEM-EUSO Science Objectives}
\maketitle

\begin{abstract}
JEM-EUSO, on board of the Japanese Exploration Module of the International Space Station, is being proposed as the first space observatory devoted to UHECR. Its privileged position at 430 km above the Earth surface, combined with a large field of view, innovative optics and a high efficiency focal surface, results in an unprecedented exposure which significantly surpasses that of the largest ground observatories. The large number of events expected above the GZK threshold for photo-pion production by protons will allow the directional identification of individual sources and the determination of their spectra, i.e., doing astronomy and astrophysics through the particle channel. Similar goals can be achieved in the case of light UHECR nuclei. Furthermore, the atmospheric target volume ($\sim 10^{12}$ ton) makes the possibility of neutrino observation a highlight of the mission. Other exploratory objectives include the detection of extreme energy gammas and the study of Galactic magnetic fields as well as global observations of the earth's atmosphere, including clouds, night-glows, plasma discharges, and meteors. In this contribution we will describe the scientific objectives of JEM-EUSO. \end{abstract}


 
\section{Introduction} \label{sec:intro}

Cosmic rays (CR) span more than 30 orders of magnitude in flux and 11 orders of magnitude in energy. At the lowest energies they originate in our Galaxy and can influence terrestrial phenomena ranging from the weather to the evolution of life. At the highest energies they may be messengers of the most extreme environments in the universe. It is this challenging extreme energy region, at the frontier of present scientific knowledge, to which the JEM-EUSO mission is devoted. JEM-EUSO is intended to address basic problems of fundamental physics and high energy astrophysics by investigating the nature and origin of extreme energy cosmic rays (EECR) which, at $E > 5 \times 10^{19}$ eV, constitute the most energetic component of the cosmic radiation. JEM-EUSO will pioneer the observation from space of EECR-induced extensive air showers (EAS), making accurate measurements of the energy, arrival direction and identity of the primary particle using a target volume far greater than which is possible from the ground. The corresponding quantitative jump in statistics will clarify the origin (sources) of the EECR, the environments traversed during production and propagation, and, possibly, on particle physics mechanisms operating at energies well beyond those achievable by man-made accelerators. Furthermore, the spectrum of scientific goals of the JEM-EUSO mission also includes as exploratory objectives the detection of high energy gamma rays and neutrinos, the study of cosmic magnetic fields, and testing relativity and quantum gravity effects at extreme energies. In parallel, all along the mission, JEM-EUSO will systematically survey atmospheric phenomena over the Earth surface. \\
 
\section{Main objectives} \label{sec:MainObjectives}

The cosmic microwave background (CMB) should lead to a suppression of the flux of protons above, roughly, $5 \times 10^{19}$ eV via photo-pion production. Much the same applies to heavier nuclei, for which the main interaction channel is photo-dissociation due to interactions with the CMB and the diffuse infrared background. The spectral signature is the well-known GZK suppression \cite{GZK}, which translates into a rather small local volume of the nearby universe which may contain the sources of the observed EECR. A GZK horizon can be defined as the distance from which $50$\% of the injected flux survives. At $6 \times10^{19}$ eV the GZK horizon is $\sim 150$ Mpc for p and Fe and much smaller for intermediate nuclei \cite{Allard2006}. Thus, the EECR sources are nearby in cosmological terms. AGASA \cite{AGASA} and HiRes \cite{HiRes} reported conflicting results: HiRes reported a spectrum close to the GZK expectation, while AGASA deduced a much harder spectrum compatible with no flux suppression. This discrepancy motivated the postulation of top-down production models and the possibility of Lorentz invariance violation at high relativistic factors. More recently, the Pierre Auger Observatory \cite{PAO}, after attaining $\sim 5 \times 10^{3}$ km$^{2}$ sr yr, measured a spectrum showing a flux suppression compatible with HiRes observations \cite{AugerSpectrum}. 
 
\begin{figure}[!h]
\centering
\includegraphics [width=0.48\textwidth]{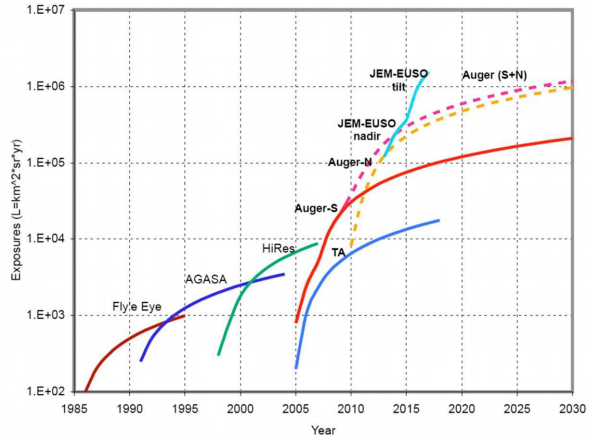}
\caption{Expected cumulative exposure of JEM-EUSO (thick curve). For comparison, the evolution of 
exposure for other past and running EECR observatories is shown.}
\label{Exposure}
\end{figure}

Although the existence of a sharp flux suppression above $4 \times 10^{19}$ eV is now firmly established, ambiguity remains as to whether this feature results from the GZK mechanism or due to other mechanisms, e.g. an intrinsic cut-off in the acceleration mechanism at the sources or a local deficit of sources. Charged particles must be contained long enough inside the source for acceleration to take place, i.e., their cyclotron radius or effective diffusion length scale must be smaller than the 
size of the acceleration region. At $10^{20}$ eV, very few astrophysical objects meet this minimum requirement \cite{HillasPlot}. Indeed, the potential candidates, e.g., GRBs, AGNs, radio galaxies and galaxy clusters bearly do so, suggesting either the possible existence of an acceleration limit at a position coincident with that of the expected GZK cut-off, or the existence of astrophysical objects or acceleration mechanisms that are unknown today. A definite answer to this fundamental question can only be obtained by measuring the CR spectrum at energies above few times $10^{20}$ eV, which requires a quantitative jump in exposure, $10^{6}$ km$^{2}$ sr yr, as only JEM-EUSO can deliver in just few years of operation (see Figure \ref{Exposure}).

\begin{figure}[!h]
\centering
\includegraphics [width=0.48\textwidth]{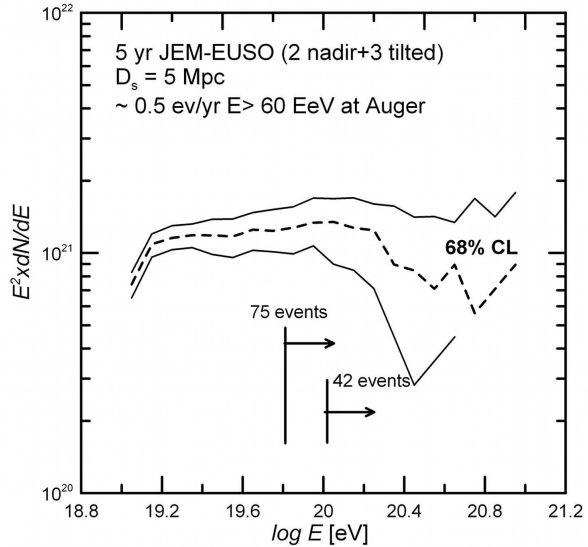}
\caption{JEM-EUSO observation of an hypothetical proton source at 5 Mpc with injected spectrum 
$\propto E^{-2}$ and a luminosity such that its flux at Auger South would be $\sim 0.5$ yr$^{-1}$ 
above $60$ EeV.}
\label{SingleSourceSpectrumICRC}
\end{figure}

Since the distribution of matter, both luminous and dark, is inhomogeneous within the GZK horizon, the EECR flux should be anisotropic above sufficiently high energies. Despite some hints of EECR anisotropy by both AGASA and HiRes, the first definitive confirmation of large-scale anisotropy was produced recently by Auger \cite{AugerAniso}. The result points to the possibility that the EECR sources are astrophysical objects whose spatial distribution is similar to the general distribution of matter at depths of some tens of Mpc. Furthermore, if several sources lie within the GZK horizon, JEM-EUSO will be the first instrument ever to have the potential to observe large clusters of events associated to individual sources, in case the primaries are protons or light nuclei, and measure their spectra (see Figure \ref{SingleSourceSpectrumICRC}).

\begin{figure}[!h]
\centering
\includegraphics [width=0.45\textwidth]{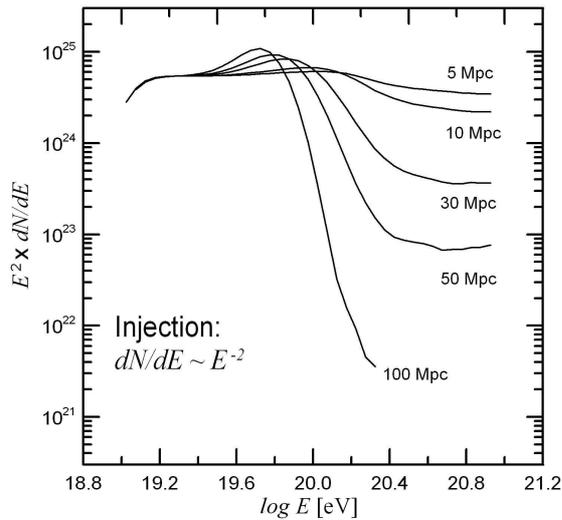}
\caption{EECR spectrum at Earth showing the strong dependence of the GZK feature on source distance. Injected spectrum is  $E^{-2}$.}
\label{dNdE}
\end{figure}

With JEM-EUSO, several tens of events are expected per source and, since the structure of the GZK feature is highly dependent on distance (see Figure \ref{dNdE}), the combination of anisotropy and spectral information can help to pinpoint the type of astrophysical objects responsible for the origin of EECRs and even, possibly, to identify individual sources.

High statistics should also allow to discover other multipole anisotropies. These could result from large scale nearby cosmic structure and/or sub-dominant components in the EECR flux (such as decay of super-heavy relics in the Galactic halo \cite{SHRHalo} or high-energy neutrino annihilation on the relic neutrino background - the Z-burst mechanism \cite{Zburst}. The latter is recognized as the unique window to the cosmic neutrino background. 

\section{Exploratory objectives}

Gamma rays at extreme energies are a natural consequence of $\pi^{0}$ production during 
EECR propagation through the CMB. A gamma-ray flux higher than expected from this secondary 
production would signify a new production mechanism, such as top-down decay/annihilation, 
or a breaking of Lorentz symmetry. The discrimination between gammas and protons at high 
energies depends on the interplay of two competing phenomena: the LPM effect, which tends 
to increase the depth of the maximum development of an EAS in the atmosphere, $X_{max}$, and pair 
production in the geomagnetic field, which moves upwards, between $1000$ and $2000$ km in height, 
the starting point of the electromagnetic cascades. The latter phenomenon depends strongly on the 
intensity of the perpendicular component of the geomagnetic field, and therefore on the impinging position 
over the Earth and incidence direction of the EECR. JEM-EUSO, with its large exposure and full 
sky coverage is ideally suited to explore these factors and fix upper limit of strong astrophysical 
significance.

\begin{figure}[!t]
\centering
\includegraphics [width=0.48\textwidth]{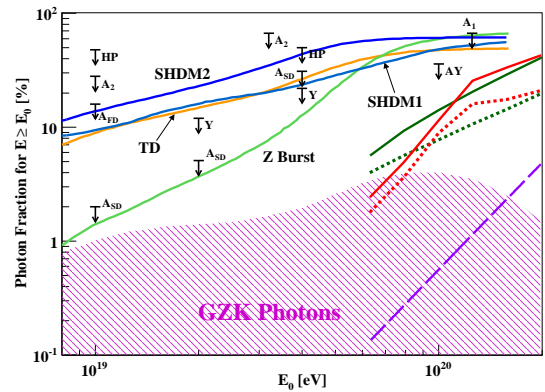}
\caption{Upper limits on the fraction of photons in the integral cosmic ray flux. Lines correspond to JEM-EUSO predictions. The dashed line corresponds to the ideal case in which it is known that there are no photons in the data. The four lines at higher photon fraction, on the other hand, correspond to the case in which it is not known whether photons exist in the data and to different assumptions on the $X_{max}$ reconstruction errors and photon fraction estimation techniques \cite{JEMEUSOGammas}. The shadow region is a prediction for the GZK  photons. Black arrows are experimental limits, HP: Haverah Park; A$_1$, A$_2$: AGASA; A$_{\textrm{FD}}$, A$_{\textrm{SD}}$: Auger; AY: AGASA-Yakutsk}
\label{GammaUpperLimit}
\end{figure}

\begin{figure}[h]
\centering
\includegraphics [width=0.45\textwidth]{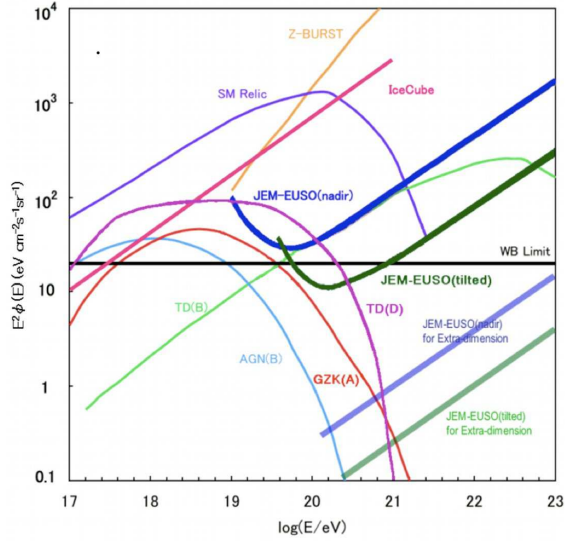}
\caption{Flux sensitivity for JEM-EUSO detecting 1 neutrino /energy decade/yr (25\% duty-cycle is assumed). Blue and green dark curves shows the case of nadir and tilted modes, respectively. Straight lines at high energies indicate the case of extra-dimension model. For IceCube (red line), 2.3 events/energy-decade/10yrs is assumed. The black line denotes the Waxman-Bahcall limit.}
\label{NeutrinoLimits}
\end{figure}

\begin{figure}[h]
\centering
\includegraphics [width=0.48\textwidth]{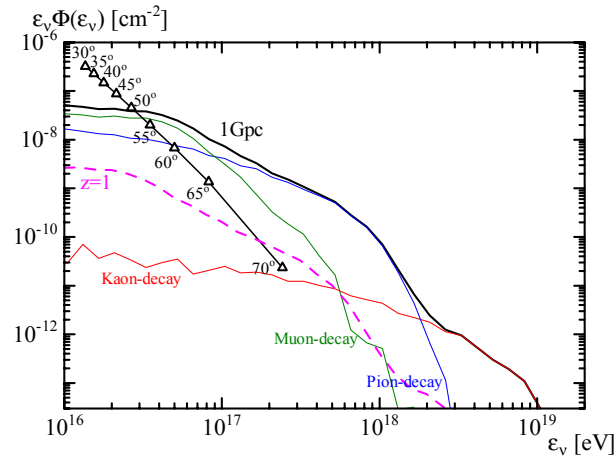}
\caption{Expected flux of $\nu_{\tau}$ from GRBs, and JEM-EUSO sensitivity for different tilting angles.}
\label{NeutrinosGRBs}
\end{figure}

During proton propagation through the CMBR, neutrinos are produced in vast amounts. These are the so 
called cosmogenic neutrinos. They are produced at two different characteristic energy ranges depending 
on whether they originate from the decay of charged pions or from neutron decay. These neutrinos 
constitute a guaranteed flux at Earth and their flux and spectrum contains extremely valuable information on the redshift evolution of the sources. Besides the cosmogenic flux there may also be contributions from hadronic interactions at the acceleration sites and from top-down processes. 
JEM-EUSO can detect neutrinos either through their LPM dominated EAS evolving deep in the 
atmosphere or, in the case of bursts of upward going neutrinos interacting inside the outermost 
layers of the crust, through direct Cherenkov. The latter scenario is expected from nearby GRB. 
Figure \ref{NeutrinoLimits} shows the flux sensitivity of JEM-EUSO for several neutrino production 
models for both, nadir and tilted mode operation of the telescope. Figure \ref{NeutrinosGRBs}, shows 
the expected flux of $\nu_{\tau}$ from GRB as function of energy, superimposed to the expected 
JEM-EUSO sensitivity \cite{JemEusoNeutrinoSensitivity} for different tilting angles. 
The discovery of ultra-high energy neutrinos beyond $10^{20}$ eV has profound implications on our
understanding of production mechanisms, since they require protons of more than $10^{21}$ eV
at the source, which is already the expected upper limit for non-relativistic or relativistic shock-wave acceleration. Higher energy neutrinos should originate either by top-down mechanisms 
or by less understood boton-up channels, like exotic plasma phenomena or unipolar induction in 
extreme environments.

Furthermore, the $\nu$ cross-section is uncertain and highly model-dependent. Extra-dimension 
models \cite{NuExtraDim} in which the Universe is supposed to consist of ten or eleven dimensions 
are among the favored models to unify the quantum mechanics and gravitation theory. In these models, 
the predicted neutrino cross-section is $10^{2}$ times larger than the Standard model prediction. 
Under these conditions, JEM-EUSO should observe $100$s of $\nu$ events, which would immediately validate experimentally low-scale unification. In addition, the ratio of horizontal to upward $\nu$-originated EAS gives a quantitative estimation of $\nu$ cross-section around $10^{14}$ eV center of mass energies \cite{NuCrossSectionEstimate}.

Additionally, a stringent test of relativity could be made from high multiplicity sources at known distances. If the GZK steepening functions consistently deviate at some directions in the sky external fields, like vector fields, might be emerging which are not unidirectionally Lorentz Invariant. On the other hand, the proof of non-vector fields would verify Lorentz Invariance at EHE \cite{N_Myu}. 

\begin{figure}[h]
\centering
\includegraphics [width=0.48\textwidth]{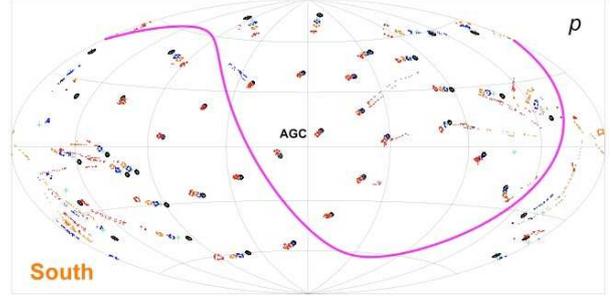}
\caption{Deformation of the point spread function of individual sources as a function of energy and
location on the sky for a certain possible realization of the Galactic magnetic field. Black corresponds to the highest energy, $10^{20}$ eV and red to the lowest, $10^{19.4}$ eV. Different realizations produce distinctive patterns \cite{polarization}.}
\label{GMFdetermination}
\end{figure}

Once point sources have been identified, the distortions suffered by the point spread function of the 
source images as a function of position in the sky and energy of the incoming particles can be used 
to deconvolve the structure of cosmic magnetic fields (see Figure \ref{GMFdetermination}).

Taking advantage of its atmospheric sounding capabilities, JEM-EUSO will also observe atmospheric luminous phenomena such as lightning, nightglow, and meteors which are expected to have manifestations in the UV band \cite{atmospherics}.

\section{Conclusions}
 
The EECR window to the universe is about to be opened, and it will be realized from space. 
JEM-EUSO is the pioneer of this new observational era.
 
%


\begin{thebibliography}{99}


\bibitem{GZK} K. Greisen 1966, Phys. Lett. 16, 148. G. T Zatsepin, V. A. KuzÕmin 1966, JETP Phys. Lett. 4, 78.

\bibitem{Allard2006} D. Allard et al. 2006, JCAP 09 005

\bibitem{AGASA} http://www-akeno.icrr.u-tokyo.ac.jp/AGASA/

\bibitem{HiRes} http://www.cosmic-ray.org/

\bibitem{PAO} http://www.auger.org

\bibitem{AugerSpectrum} The Pierre Ager Collaboration 2008, PRL 101, 061101.

\bibitem{HillasPlot} Hillas A. M. 1984, Ann. Rev. Astron. Astrophys. 22, 425.

\bibitem{AugerAniso} The Pierre Auger Collaboration 2007, Science 318, 938.


\bibitem{SHRHalo} Berezinsky V., Kachelriess M. and Vilenkin A. 1997, Phys Rev Letters 79 4302. Medina-Tanco, G. and Watson, A. 1999, Astropart. Phys. ,12, 25.

\bibitem{Zburst} Weiler T. 1982, Phys. Rev. Lett. 49, 234. Weiler T. 1984, Astrophys. J. 285, 495.

\bibitem{PSF} Medina Tanco for the JEM-EUSO Collaboration 2009, Proc. 31st ICRC, Poland, ID 138.

\bibitem{JEMEUSOGammas} D.~Supanitsky et al. for the JEM-EUSO Collaboration, Proc. 31st ICRC, Poland, ID 1423.  









\bibitem{JemEusoNeutrinoSensitivity} Asano K ad Nagataki S. 2006, Astrophys J. 640, L9.

\bibitem{NuExtraDim} Ancordoqui L. A., Feng J. L. and Goldberg H. 2002, Phys. Lett. B 525, 302.

\bibitem{NuCrossSectionEstimate} Palimares-Ruiz S, Irimia A. and Weiler T. J. 2006, Phys. Rev. D 73, 
083003.

\bibitem{N_Myu} V. A. Kostelecky, Phys.Rev. D69 (2004) 105009

\bibitem{polarization} Medina-Tanco, G., Teshima, M. and Takeda, M. 2001, 29th ICRC, Tsukuba, 747.

\bibitem{atmospherics} Horinouchi T., Nakamura T., Kosaka J. 2007, J. Geophys. Res. Lett., 29. 

\end{thebibliography}
\end{document}